\documentclass[article,12pt]{elsarticle}     

\usepackage{graphicx,epsf,color}
\usepackage{amssymb}

\begin{document}

\begin{frontmatter}



\title{Condensate of $\mu$-Bose gas as a model of dark matter}



\author{A.M. Gavrilik\footnote{E-mail: omgavr@bitp.kiev.ua}, I.I. Kachurik,
M.V. Khelashvili and A.V. Nazarenko}

\address{Bogolyubov Institute for Theoretical Physics - 14-b, Metrolohichna str.
 Kiev, 03143, Ukraine}



\begin{abstract}
Though very popular, Bose-Einstein condensate models of dark matter
have some difficulties. Here we propose the so-called $\mu$-Bose gas
model ($\mu$-BGM) as a model of dark matter, able to treat weak
points. Within $\mu$-BGM, the $\mu$-dependence of thermodynamics
arises through the respective $\mu$-calculus (it generalizes usual
differential calculus) and enters the partition function, total
number of particles, internal energy, etc. We study thermodynamic
geometry of the $\mu$-BGM and find singular behavior of (scalar)
curvature, confirming Bose-like condensation. The critical
temperature of condensation $T^{(\mu)}_c$ for $\mu\neq 0$ is higher
than the boson $T_c$. We find other important virtues of
$\mu$-thermodynamics versus usual bosons and conclude: the
condensate of $\mu$-Bose gas can serve as (an effective) model of
{galactic-halos dark matter}.
\end{abstract}

\begin{keyword}
boson condensate \sep deformed Bose gas model \sep nonstandard
statistics \sep critical temperature \sep thermodynamic geometry
\sep dark matter halo
\PACS 05.70.Ce \sep 05.30.Jp \sep 03.75.Hh \sep 95.35.+d
\end{keyword}

\end{frontmatter}











\newpage

\section{Introduction}
Among the approaches to model dark matter, those exploiting
Bose-Einstein condensate (BEC) are very popular, see
e.g.~\cite{Sin,Boehmer,Harko,Arbey,Hu,Kain} and the
review~\cite{review}. They share nice features of cold dark matter,
show a number of advantages, but also encounter their own
difficulties, e.g. the problem of gravitational
collapse~\cite{Guzman}, overestimated dark halo mass~\cite{Harko}
etc.

As the true nature of dark matter constituents is unknown, many
exotic candidates were considered, e.g. axionic~\cite{axion} or even
stringy ones~\cite{stringy}. In some papers, the authors exploit
certain models of nonstandard thermostatistics with aim to describe
basic objects of quantum cosmology~\cite{Tavayef}, the physics of
dark matter~\cite{Infinite,Dil} or black
holes~\cite{Strominger,Ng,Zare}. It is worth to examine various
nontrivial models of nonstandard thermostatistics as possible
candidates for modeling, at least effectively, main properties of
dark matter in order to choose most adequate one.

In this paper we explore the $\mu$-Bose gas model as a possible
model of dark matter, and come to important facts. The $\mu$-Bose
gas model was first  proposed in~\cite{GR-intercepts,GM-correl}
where the correlation functions intercepts of 2nd and higher order
were derived. The study of $\mu$-Bose gas thermodynamics started
in~\cite{RGK} and used the special so-called $\mu$-calculus. That
allowed to explore basic quantities, e.g. $\mu$-analogs of
elementary and special functions.

There are diverse deformed Bose gas models, see
e.g.~\cite{GR-intercepts,GM-correl,Martin91,Manko,
Chaichian,Monteiro,ShuChen02,AdGavr,GavrSigma,AlginFibOsc,ScarfoneSwamy09,GR-12}.
As usual, the deformations are based on respective deformed
oscillator (DO) models such as
$q$-oscillators~\cite{ArikCoon76,Biedenharn} or the 2-parameter
$p,q$-deformed (or Fibonacci) oscillators~\cite{Chakrabarti91}. A
plenty of nonstandard one-parameter DOs exist~\cite{Plethora08},
along with polynomially deformed ones~\cite{Polynomially}.
 Among the so-called quasi-Fibonacci oscillators~\cite{Kachurik} we find the
$\mu$-oscillator~\cite{Jann}.  Note that unlike DOs of polynomial
type, the $\mu$-oscillator (and $\mu$-bosons) belong to the less
studied class of rational type DOs. The DO models often possess
unusual properties, e.g. energy level degeneracies, nontrivial
recurrent relations for energy spectra etc. Nontrivial features of
DOs motivate their application in diverse fields of quantum physics.

Physical meaning of deformation parameter(s) of deformed model
depends on its specific application to physical system. Say, when
the model of ideal gas of {\it deformed} bosons is applied to
compute the intercepts of the momentum correlation
functions~\cite{GR-intercepts,AdGavr,GavrSigma}, one effectively
takes into account the non-zero proper volume of
particles~\cite{av95}, or their internal structure, or
compositeness~\cite{GKM2011,GM2012}. The experimental data on the
two-pion correlation-function intercepts unravel~\cite{Abelev}
non-Bose type behavior of pions, and the use of deformed BGM has
shown its efficiency~\cite{GavrSigma,AGP,GM_NP}. There exists an
application of $q$-Bose gas setup to description of the phonon
spectrum of $^4$He, and the agreement with experiment is
obvious~\cite{Monteiro}.

In the $\mu$-BGM and other deformed analogs of Bose gas model,
quantum statistical interaction gets
modified~\cite{GR-12,AlginSenay}.  Moreover, deformation can also
absorb~\cite{ScarfoneSwamy09} an interaction present in the
initially non-deformed system.

About the plan of our paper. In Sec.~2 we give a setup of the
$\mu$-deformed Bose gas model and of $\mu$-calculus. Thermodynamical
quantities are considered in Sec.~3: the total number of particles
is given explicitly and from it -- the partition function (all with
explicit $\mu$-dependence). In Sec.~4, within geometric approach to
thermodynamics (see e.g.~\cite{Ruppeiner, Janyszek,Quevedo,Ubriaco})
we confirm the existence of Bose-like condensation in $\mu$-Bose
gas. Critical temperature of condensation and its dependence on the
deformation parameter $\mu$ are studied. Other aspects or
thermodynamical functions, useful for the application, are
considered in the next section. Discussion of most important
features and virtues of $\mu$-Bose gas enabling its application to
model dark matter, and the concluding remarks, are given in the
final section of the paper.

\section{Deformed analogs of Bose gas model}

Like in other works on deformed oscillators, see e.g.~\cite{Monteiro,ScarfoneSwamy09,AlginFibOsc},
we deal in fact with the (system of) deformed bosons. The important virtues of such deformation is
its ability to provide effective account of interaction between particles, their non-zero volume,
their inner (composite) structure etc.

The $\mu$-deformed BGM associated with $\mu$-oscillator~\cite{Jann}
was introduced in~\cite{GR-intercepts, GM-correl}. Therein and in
this paper the thermal average of the operator $\mathcal{O}$ is
determined by the formula   
\begin{equation}\label{eq.2}
\langle \mathcal{O} \rangle=\frac{Tr(\mathcal{O}e^{-\beta H})}{Z},
\end{equation}
$Z$ being the grand canonical partition function. Its logarithm is
\begin{equation}\label{eq.3}
\ln Z=-\sum_i\ln(1-ze^{-\beta\varepsilon_i})
\end{equation}
with the fugacity $z=e^{\beta\widetilde{\mu}}$\, ($\widetilde{\mu}$
is chemical potential). The familiar formula
\begin{equation}\label{eq.4}
N=z\frac{d}{dz}\ln Z
\end{equation}
for the number of particles will be modified (deformed), see below.

To study the $\mu$-BGM as the model for the system of deformed bosons, we use
the Hamiltonian with chemical potential $\widetilde{\mu}$
\begin{equation}\label{eq.1}
H=\sum_i(\varepsilon_i-\widetilde{\mu})N_i\, .
\end{equation}
Here $\varepsilon_i$ is kinetic energy of particle in the state
"$i$" and $N_i$ the particle number (occupation number) operator
corresponding to state "$i$".

\underline{\it Elements of $\mu$-calculus}.
 To develop the $\mu$-analog of BGM, we extend (deform) the notion of derivative.
 So-called $\mu$-derivative, introduced in~\cite{RGK}, differs from the known
 Jackson or $q$-derivative~\cite{Kac} and its $p,q$-extension (used in~\cite{GR-12}).
 The easiest way to define the $\mu$-extension is to apply the rule
\begin{equation}\label{eq.10}
\mathcal{D}^{(\mu)}_x x^n\!=\![n]_{\mu}x^{n\!-\!1},  \quad
[n]_{\mu}\!\equiv\!\frac{n}{1+\mu n} \quad \mbox{($\mu$-bracket)}\,
\end{equation}
so that the $\mu$-derivative involves $\mu$-bracket from the work on $\mu$-oscillator~\cite{Jann}.
If $\mu\!\rightarrow\! 0$, then $[n]_{\mu}\to n$ and the $\mu$-extension
$\mathcal{D}^{(\mu)}_x$ reduces to ordinary $d/dx$. 

Formula for $\mu$-derivative acting on monomials $x^m$ is enough for us in this work,
while the general rule for $\mu$-derivative action on a function $f(x)$ is
\begin{equation}\label{eq.11}
\mathcal{D}^{(\mu)}_xf(x)=\int^1_0dtf'_x(t^{\mu}x),  \qquad
f'_x(t^{\mu}x)=\frac{d f(t^{\mu}x)}{d x}.
\end{equation}
Clearly, formula (\ref{eq.10}) stems from this general definition.
Note that the inverse $\bigl({\mathcal{D}^{(\mu)}_x}\bigr)^{-1}$ of
the $\mu$-derivative $D^{(\mu)}_x$ in (\ref{eq.10}) and
(\ref{eq.11}) is also known (we omit it).

For $k$th power of $\mu$-derivative acting on $x^n$ we have
\begin{equation}\label{eq.12}
(\mathcal{D}^{(\mu)}_x)^kx^n=\frac{[n]_{\mu}!}{[n-k]_{\mu}!}x^{n-k},
\qquad [n]_{\mu}!\equiv\frac{n!}{(n;\mu)}
\end{equation}
where $(n;\mu)\equiv(1+\mu)(1+2\mu)...(1+n\mu)$.

There exist certain $q,{\mu}$- or $(p,q;\mu)$-deformed extensions
of $\mu$-derivative:   
instead of $(d/dx)f(t^{\mu}x)$ in (\ref{eq.11}) take
$\mathcal{D}^{q}_xf(t^{\mu}x)$ or $\mathcal{D}^{(p,q)}_xf(t^{\mu}x)$.
The two extensions correspond to the quasi-Fibonacci $(q;\mu)$-DO or
$(p,q;\mu)$-DO in~\cite{Kachurik}.

 So, to develop the $\mu$-Bose gas thermodynamics we replace,
 where necessary, the usual $d/dz$ with $\mu$-derivative $D^{(\mu)}_z$.
 Due to the $\mu$-derivative, basic parameter $\mu$ enters the treatment:
 the system gets $\mu$-deformed.
    For $\mu \ll 1$, the usual and $\mu$-deformed
 derivatives of a function have similar behavior, that
  is easily seen by acting with $\mu$-derivative and usual one
 on the monomial, logarithmic, exponential function, etc.
  Such property of $\mu$-derivative can justify  
  its use in developing thermodynamics of $\mu$-Bose gas.

 The $\mu$-bracket $[n]_{\mu}$ and $\mu$-factorial
$[n]_{\mu}!$, see (\ref{eq.10}), (\ref{eq.12}), generate
$\mu$-deformed analogs~\cite{RGK} of familiar functions:
 $\mu$-exponential $\exp_{\mu}(x)$, $\mu$-logarithm $\ln_{\mu}(x)$
 (with $\mu$-numbers $[n]_{\mu}$ and $\mu$-factorial
$[n]_{\mu}!=[n]_{\mu}[n-1]_{\mu}...[2]_{\mu}[1]_{\mu}$).
 New special functions e.g. $\mu$-polylogarithms
 do also appear, see~\cite{RGK} and below.

Formula for $D^{(\mu)}_x$ operating on general product $f(x)\cdot g(x)$ is also known.

\section{Thermodynamics of $\mu$-Bose gas model}

Thermodynamics of $\mu$-BGM is based on $\mu$-calculus. We consider
the gas of non-relativistic particles {for the regimes
\cite{Pathria} of both high and low temperatures\footnote{These
regimes are respectively given by the inequalities
$\frac{v}{\lambda^3}\gg 1$ and $\frac{v}{\lambda^3}\ll 1$ which
involve the specific volume $v=V/N$ and thermal wavelength
$\lambda$, see (\ref{eq.22}) and below.}}.


\underline{\it Total number of particles}.
 The usual relation for total number of Bose gas particles is given in (\ref{eq.4}).
 For $\mu$-BGM, the total number of particles $N\equiv N^{(\mu)}$ is defined as
$N^{(\mu)}=z\mathcal{D}^{(\mu)}_z\ln Z$ using $\mu$-derivative
$\mathcal{D}^{(\mu)}$ from (\ref{eq.10}). Applying it, for $\mu\geq
0$, to the ($\log$ of) partition function in (\ref{eq.3}) we get
\begin{equation}\label{eq.16}
N^{(\mu)}\!=\!
 \sum_i\!\sum_{n=1}^{\infty}\frac{[n]_{\mu}}{n}e^{-\beta\varepsilon_in}z^n\, .
\end{equation}
Set $0\leq|ze^{-\beta\varepsilon_i}|<1$ in (\ref{eq.16}).
 For non-relativistic particles the energy $\varepsilon_i$
 is\footnote{Similarly to refs.~\cite{AlginFibOsc,GR-12} and others,
 we initially take the particle kinetic energy as that of non-relativistic {\it free particle}.
 But, the very particles in the model are not the usual bosons, because of the deformation
 of thermodynamics that uses $\mu$-calculus. As result, all thermodynamical quantities
 including the mean kinetic energy of particle become dependent on the
 deformation parameter $\mu$.}
\begin{equation}\label{eq.17}
\varepsilon_i=\frac{\overrightarrow{p}_i\overrightarrow{p}_i}{2m}
=\frac{p_i^2}{2m}\, ,
\end{equation}
involving 3-momentum $\overrightarrow{p}_i$ of particle of mass $m$
in the $i$-th state.

 At $z\rightarrow 1$ the summand in (\ref{eq.16}) diverges if $p_i=0, i=0$.
 Suppose that the $i=0$ ground state admits macroscopically large occupation number.
 For $z\neq 1$ we as well separate the term with $p_i=0$ from the remaining sum:
\begin{equation}\label{eq.18}
N^{(\mu)}={\sum_i} ' \sum_{n=1}^{\infty}\frac{[n]_{\mu}}{n}(e^{-\beta\varepsilon_i})^nz^n +
\sum_{n=1}^{\infty}\frac{[n]_{\mu}}{n}z^n.
\end{equation}
The symbol ${\sum_i}'$ means that the $i\!=\!0$ term is dropped from the sum.
For large volume $V$ and large $N$ the spectrum of single-particle states is almost continuous
so we replace the sum $\sum_i\rightarrow \frac{V}{(2\pi \hbar)^3}\int d^3p$
(the ground state, $p_0$, does not contribute to the integral, so it starts from zero).

Further calculation of $N^{(\mu)}$ proceeds (in $d=3$) similarly to
the case of $p,q$-Bose gas~\cite{GR-12} (see also ref.~\cite{RGK}
for details), and the final result reads:
\begin{equation}\label{eq.22}
N^{(\mu)}=\frac{V}{\lambda^3}g_{3/2}^{(\mu)}(z)+g_0^{(\mu)}(z),
\qquad  g_0^{(\mu)}(z)=N_0^{(\mu)} \, .
\end{equation}
Here $\lambda\!=\!\sqrt{\frac{2\pi\hbar^2}{mkT}}$ is the thermal  wavelength; $g_0^{(\mu)}(z)$
and $g_{3/2}^{(\mu)}(z)$ are $\mu$-analogs of polylogarithm $g_l(z)\!=\!\sum_{n=1}^{\infty}z^n/n^l$,
or $\mu$-polylogarithms, defined as
\begin{equation}\label{eq.23}
g_{l}^{(\mu)}(z)=\sum_{n=1}^{\infty}\frac{[n]_{\mu}}{n^{l+1}}z^n.
\end{equation}
For real $\mu>0$ the convergence properties are not spoiled  (like for the usual $g_l(z)$,
there should be $|z|<1$).  If $\mu\rightarrow 0$, we recover polylogarithm $g_l(z)$.

For further needs, it is convenient to rewrite the expression
(\ref{eq.22}) in terms of volume per particle $v\equiv V/N^{(\mu)}$.

\underline{\it Deformed grand partition function}.
 In $\mu$-BGM, we use the relations between thermodynamic functions
 similar to those of usual Bose gas thermodynamics, but in our case all the
thermodynamic functions like the partition function etc., become
$\mu$-dependent.

The deformed partition function $\ln Z^{(\mu)}$ satisfies
\begin{equation}\label{eq.25}
N^{(\mu)}=z\frac{d}{dz}\ln Z^{(\mu)}
\qquad  {\rm or } \qquad
\ln Z^{(\mu)}=\Bigl(z\frac{d}{dz}\Bigr)^{-1}N^{(\mu)}.
\end{equation}
To apply $\bigl(z\frac{d}{dz}\bigr)^{-1}$, we use the
property $f\Bigl(z\frac{d}{dz}\Bigr)z^k=f(k)z^k$ for a function
$f(X)$ which admits power series expansion. From (\ref{eq.25}),
(\ref{eq.22}), (\ref{eq.23}) we infer
\begin{equation}\label{eq.28}
\ln Z^{(\mu)}=\frac{V}{\lambda^3}
 \sum_{n=1}^{\infty}\frac{[n]_{\mu}}{n^{5/2}}(n)^{-1}z^n
 +\sum_{n=1}^{\infty}\frac{[n]_{\mu}}{n}(n)^{-1}z^n .
\end{equation}
In a more compact form,
\begin{equation}\label{eq.29}
Z^{(\mu)}(z,T,V)=\exp\biggl(\frac{V}{\lambda^3}g^{(\mu)}_{5/2}(z)+g^{(\mu)}_1(z)\!\biggr).
\end{equation}
Formulas (\ref{eq.28})-(\ref{eq.29}) provide the $\mu$-partition function
and play basic role: from these 
we can derive other thermodynamical functions and relations.

\section{Geometric approach to $\mu$-Bose gas model}

We study the $\mu$-thermodynamics using thermodynamic geometry
in the space of two parameters $\beta, \gamma$, where $\gamma=-\beta\tilde{\mu}$ and
$\tilde{\mu}$ is chemical potential.
 The (scalar) curvature in the thermodynamic parameters space
provides the efficient and elegant tools for exploring
thermodynamical properties of the system under
study~\cite{Ruppeiner,Janyszek,Ubriaco}. This geometric construction
is formed, see below, by the derivatives of a thermodynamic
potential that determines a surface in the space of thermodynamic
parameters. The curvature reveals the extremal points of this
surface, identified~\cite{Quevedo} with phase transitions: its
singularity gives a sufficient condition for the existence of the
phase transition point(s). Moreover, the curvature is nothing but a
thermodynamic measure of interaction within the system, and its sign
indicates attractive or repulsive character of this interaction.

The components of (symmetric) metric in the Fisher-Rao
representation are
\begin{equation}
G_{\beta\beta}=\frac{\partial^2\ln Z^{(\mu)}}{\partial\beta^2},\quad
G_{\beta\gamma}=\frac{\partial^2\ln Z^{(\mu)}}{\partial\gamma\,\partial\beta},\quad
G_{\gamma\gamma}=\frac{\partial^2\ln Z^{(\mu)}}{\partial\gamma^2}.
\label{met}
\end{equation}
Using the thermodynamic relations, we also have
$$
G_{\beta\beta}=-\left(\frac{\partial U}{\partial \beta}\right)_{\gamma},\quad
G_{\beta\gamma}=-\left(\frac{\partial N}{\partial\beta}\right)_{\gamma},\quad
G_{\gamma\gamma}=-\left(\frac{\partial N}{\partial \gamma}\right)_{\beta}.
$$
Remembering (\ref{eq.29}), we arrive at the metric (given by
$\mu$-polylogarithms):
\begin{eqnarray}
G_{\beta\beta}&=&\frac{15}{4}\frac{V}{\lambda^3\beta^2}\,g^{(\mu)}_{\frac{5}{2}}(z),
 \hspace{11mm}
G_{\beta\gamma}=\frac{3}{2}\frac{V}{\lambda^3\beta}\,g^{(\mu)}_{\frac{3}{2}}(z),\\
G_{\gamma\gamma}&=&\frac{V}{\lambda^3}\,g^{(\mu)}_{\frac{1}{2}}(z)+g^{(\mu)}_{-1}(z).
\end{eqnarray}
 The determinant of the metric $g\equiv\det|G_{ij}|$ results as
 (let $\beta\leftrightarrow 1,\, \gamma\leftrightarrow 2$)
\begin{equation}
g\!=\!\frac{3V}{4\lambda^3\beta^2}\left(5g^{(\mu)}_{\frac{5}{2}}(z)
g^{(\mu)}_{-1}(z)+\frac{V}{\lambda^3}\left(5g^{(\mu)}_{\frac{5}{2}}(z)
g^{(\mu)}_{\frac{1}{2}}(z)\!-\!3g^{(\mu)}_{\frac{3}{2}}(z)
g^{(\mu)}_{\frac{3}{2}}(z)\right)\right)
\end{equation}
and the inverse metric is\ 
$G^{11}=G_{22}/g$, $G^{12}=-G_{12}/g$, $G^{22}=G_{11}/g$.

As the metric components are given
by the derivatives of partition function, the Christoffel symbols
and the Riemann tensor are found as
\begin{eqnarray}
\Gamma_{\lambda\sigma\nu}&=&\frac{1}{2}\bigl(\ln Z^{(\mu)}\bigr)_{, \lambda\sigma\nu},\\
R_{\lambda\sigma\nu\rho}&\equiv&
G^{\kappa\tau}\left(\Gamma_{\kappa\lambda\rho}\Gamma_{\tau\sigma\nu}
- \Gamma_{\kappa\lambda\nu}\Gamma_{\tau\sigma\rho} \right).
\end{eqnarray}
Calculation of the Christoffel symbols yields
\begin{eqnarray}
\Gamma_{\beta\beta\beta}&=& -\frac{105}{16}\frac{V}{\lambda^3\beta^3}\,g^{(\mu)}_{\frac{5}{2}}(z),\\
\Gamma_{\beta\beta\gamma}&=&\Gamma_{\beta\gamma\beta}=\Gamma_{\gamma\beta\beta}
=-\frac{15}{8}\frac{V}{\lambda^3\beta^2}\,g^{(\mu)}_{\frac{3}{2}}(z),\\
\Gamma_{\beta\gamma\gamma}&=&\Gamma_{\gamma\beta\gamma}=\Gamma_{\gamma\gamma\beta}
=-\frac{3}{4}\frac{V}{\lambda^3\beta}\,g^{(\mu)}_{\frac{1}{2}}(z),\\
\Gamma_{\gamma\gamma\gamma}&=&-\frac{1}{2}\left(\frac{V}{\lambda^3}\,g^{(\mu)}_{-\frac{1}{2}}(z) +
g^{(\mu)}_{-2}(z) \right).
\end{eqnarray}
Since the scalar curvature in $2$-dimensional space is determined by
one component of Riemann tensor $R=2R_{\beta\gamma\beta\gamma}/g$, we obtain our main
result:
\begin{equation}
R=\frac52\frac{ 5g^{(\mu)}_{\frac{3}{2}} g^{(\mu)}_{\frac{3}{2}}
g^{(\mu)}_{-1} -7g^{(\mu)}_{\frac{5}{2}} g^{(\mu)}_{\frac{1}{2}}
g^{(\mu)}_{-1} + 2g^{(\mu)}_{\frac{3}{2}} g^{(\mu)}_{\frac{5}{2}}
g^{(\mu)}_{-2} +\frac{V}{\lambda^3} \, {\cal W}}
{\left(5g^{(\mu)}_{\frac{5}{2}} g^{(\mu)}_{-1} +
\frac{V}{\lambda^3}\left(5g^{(\mu)}_{\frac{5}{2}}
g^{(\mu)}_{\frac{1}{2}} - 3g^{(\mu)}_{\frac{3}{2}}
g^{(\mu)}_{\frac{3}{2}}\right)\right)^2}\, ,
\end{equation}
$$
{\cal W}\equiv2g^{(\mu)}_{\frac{3}{2}} g^{(\mu)}_{\frac{5}{2}}
g^{(\mu)}_{-\frac{1}{2}} -4g^{(\mu)}_{\frac{5}{2}}
g^{(\mu)}_{\frac{1}{2}} g^{(\mu)}_{\frac{1}{2}}
+2g^{(\mu)}_{\frac{3}{2}} g^{(\mu)}_{\frac{3}{2}}
g^{(\mu)}_{\frac{1}{2}}.
$$

The $\mu$-polylogarithm $g^{(\mu)}_l(z)$ involved in $R$ is singular
at $z\to 1$, for $l \le 1$ in the case $\mu = 0$ and for $l \le 0$
when $\mu \ne 0$. Curvature $R(z)$ in isothermal process has
characteristic properties as a function of
fugacity\footnote{Positive sign of the curvature corresponds to
attraction among particles, and the magnitude of curvature is
growing function of $\mu$. That is natural since the parameter $\mu$
of deformation provides effective account of quantum-statistical
inter-particle interactions.}.
 In the case $v/\lambda^3 \ll 1$, $R(z)$ has no singularities as seen in
Fig.~1 (left panel).

If $v/\lambda^3$ is sufficiently large (see Fig~1, right panel) to
neglect the terms which do not  contain this factor, the curvature
$R(z)$ is singular at $z\to 1$.    We conclude that, in the latter
situation, the system undergoes phase    transition, and hence
Bose-like condensation takes place. That is, the $\mu$-Bose gas
model {\it satisfies basic necessary property}.

\begin{figure}
\begin{center}
\includegraphics[width=0.45\linewidth]{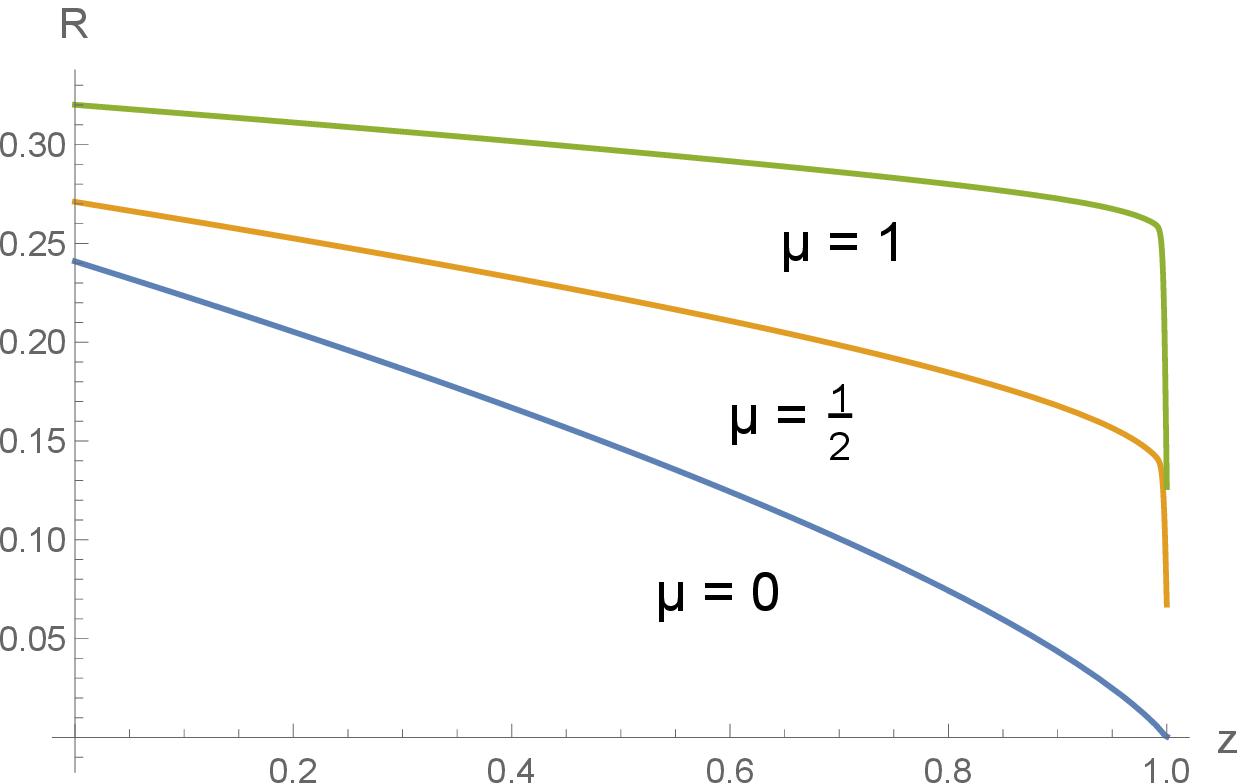}\qquad
\includegraphics[width=0.45\linewidth]{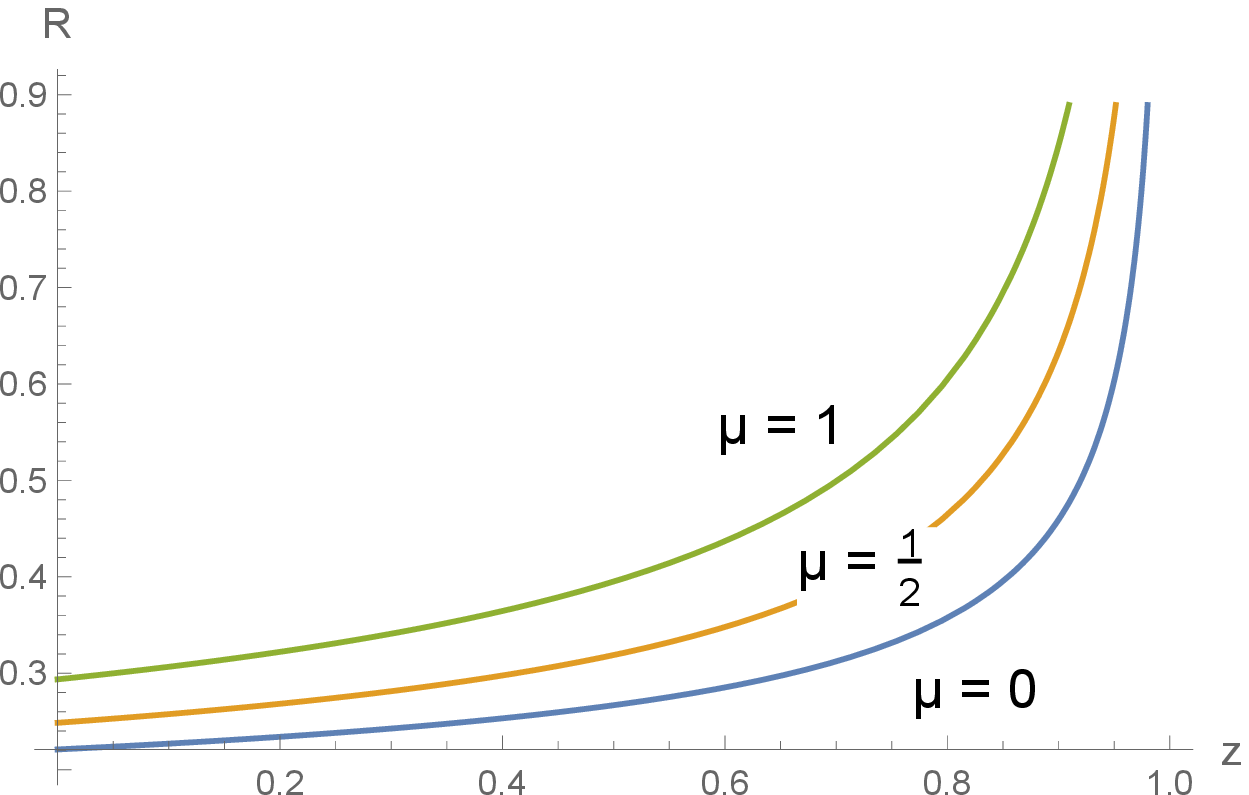}
\end{center}
\vspace*{-3mm} \caption{Scalar curvature $R(z)$ in isothermal
process $\beta={\rm const}$ for various values of deformation
parameter $\mu = 0, \frac{1}{2}, 1$. Left panel: $v/\lambda^3 \ll 1
$. Right panel: $v/\lambda^3 \gg 1$.}
\end{figure}

\section{Critical temperature of condensation}

For low temperatures and high density we can obtain~\cite{RGK}, like
for $p,q$-Bose gas \cite{GR-12}, the critical temperature
$T^{(\mu)}_c$ of condensation in the considered $\mu$-BGM.
At vanishing $N_0^{(\mu)}$ in (\ref{eq.22}),
the critical temperature $T_c^{(\mu)}$ of $\mu$-Bose gas is
determined by the equation $\lambda^3/{v}=g^{(\mu)}_{3/2}(1)$ that
gives \cite{RGK}
\begin{equation}\label{eq.37}
T_c^{(\mu)}=\frac{2\pi\hbar^2/mk}{\bigl(vg^{(\mu)}_{3/2}(1)\bigr)^{2/3}}
\quad  {\rm and} \quad
\frac{T_c^{(\mu)}}{T_c}=\Biggl(\frac{2.61}{g^{(\mu)}_{3/2}(1)}\Biggr)^{2/3}\, ,
\end{equation}
the latter being the ratio of $\mu$-critical  
 $T_c^{(\mu)}$ to the critical $T_c$ of Bose gas~\cite{Pathria}.
As is seen, the ratio $T_c^{(\mu)}/{T_c}$
has an important feature: the stronger is deformation (measured by
$\mu$) the higher is $T_c^{(\mu)}$. \ Say, for $\mu=0.06$ we have
$T_c^{(\mu)}\simeq 1.22 \cdot T_c$.
 If $\mu\!\rightarrow\!0$ (no-deformation limit), the ratio is $T_c^{(\mu)}/{T_c}=1$,
 i.e, the $\mu$-critical temperature tends to usual one,
$T_c^{(\mu)}\rightarrow T_c$ (a kind of consistency).
   The existence of condensate of $\mu$-bosons is {\it the crucial
 property} for the use of $\mu$-BGM in the   
 modeling of dark matter.

Let $T$ be in the interval $0\!<\!T\!<\!T^{(0)}_c\!\leq
T^{(\mu)}_c$.   In the $\mu$-deformed case we have
$\frac{U^{(\mu)}}{T}=\frac25 c^{(\mu)}_v= \frac35 S^{(\mu)}$ in
similarity with pure Bose case, i.e. $\frac{U}{T}=\frac25 c_v=
\frac35 S .$  From these we infer two useful relations:\
$\frac{U^{(\mu)}}{c^{(\mu)}_v}=\frac{U}{c_v}$
 and $\frac{U^{(\mu)}}{S^{(\mu)}}=\frac{U}{S}$.

\section{Other issues important for modeling dark matter}

  Above, exploiting thermodynamic geometry, we
verified main property of $\mu$-BGM needed for its ability to model
dark matter.  The appearance of Bose-like condensation as such
property, is confirmed.

As mentioned in \cite{Harko}, the dark matter surrounding dwarf
galaxies must be "{\it strongly coupled, dilute system of
particles}". In case of $\mu$-Bose gas, we emphasize that the
interaction between $\mu$-bosons (of quantum-statistical origin) is
also attractive, and due to deformation can even be stronger than
that for pure bosons. To witness, compare the two 2nd virial
coefficients: $\mu$-dependent $V_2^{(\mu)}$~\cite{RGK}, and the
 standard $V_2^{(Bose)}\!=\!2^{-5/2}$ (drop the "-" sign): 
$$
V_2^{(\mu)}-{V_2}^{Bose}= 2^{-7/2} \biggl(
\frac{[2]_{\mu}}{[1]^2_{\mu}} - 2 \biggr)
=2^{-5/2}\frac{\mu^2}{1+2\mu} > 0 .
$$
The enhanced attraction (of quantum origin) implies that the
$\mu$-bosons are "more bosonic" than usual bosons. This property is
good for providing {\it strongly coupled} system of
(quasi)particles.
We see that at large $\mu$ one has
$g^{(\mu)}_l(z)\to\mu^{-1}g^{(0)}_l(z)\equiv\mu^{-1}g_{l+1}(z)$,
where $g_l(z)$ is the polylogarithm. Then, the internal energy per
particle would not depend on $\mu$ while the total one does, due to
scale factor.

However,  unlimited growth of $\mu$-parameter (strength of
attraction) could lead to a collapse of the studied quantum system.
To prevent that, we can find some bound on the values of $\mu$, say,
requiring to forbid negative pressure.
 We take (virial expansion of) the equation of state~\cite{RGK}
 to second order i.e.
\begin{equation}\label{EoS}
\frac{P v}{k T}=1-\frac{\ \ \ [2]_{\mu}}{2^{7/2}[1]^2_{\mu}}\frac{\lambda^3}{v}\, .
\end{equation}
Impose $P=0$ or
$\frac{2^{-7/2}[2]_{\mu}}{[1]^2_{\mu}}\frac{\lambda^3}{v} = 1$ and
find critical  strength $\bar{\mu}$ of deformation:
\begin{equation}
\bar{\mu} = \kappa -1 + \sqrt{(\kappa -1)\kappa},
\hspace{10mm}
\kappa\equiv 2^{5/2} \frac{v}{\lambda^3}\, .
\end{equation}
With $\mu \leq \bar{\mu}$ we avoid the collapse. Obviously,
$\kappa\!=\!1$ means $\bar{\mu}=0$ and so $\mu=0$, that is pure Bose
case. On the other hand, the bound taken as say $\bar{\mu}=1$ with
$0 < \mu \le 1$ corresponds to the value $\kappa=\frac{4}{3}$.

Equation of state (\ref{EoS}) in its application to the dark matter
involves the temperature in the range $0<T\leq T^{(\mu)}_c$, with
non-vanishing pressure $P$ of $\mu$-BEC. At the same time, the dark
matter pressure within the ordinary BEC concept is also supposed to
be non-zero: $P=2\pi a\hbar^2\rho^2/m^3$, even at $T=0$, because of
the supposed scattering with length $a$ of bosons with the mass
density $\rho$. Since the effect of $\mu$-deformed statistics and
the scattering represent two non-identical processes, the both can
be taken together into account within an extended, unified model of
dark model in a future study.

There exists some other, "characteristic" value ${\mu}_0$ of ${\mu}$
for which the deformed entropy (see eq.~(41) and Fig.~6 in
\cite{RGK}) becomes \ $ \frac{S^{(\mu_0)}\lambda^3}{V k_B}=1$ or
just
 $\frac{S^{(\mu_0)}\lambda^3}{V}=k_B$.
With the ${\mu}_0$, we bound our interval as
$0<\mu\leq{\mu}_0$.      
At this deformation strength ${\mu}_0\simeq 1.895$, we obtain the
relation
\begin{equation}
g_{3/2}(1) = 3.3535 \ g_{3/2}^{(\mu=\mu_0)}(1)
\end{equation}
for the polylogarithm and $\mu$-polylogarithm.  Then we find that
the critical volume-per-particle (divided by cube of thermal
wavelength) in the $\mu$-BGM is related with similar quantity of the
usual Bose gas by the formula
$$ \Bigl(\frac{v^c}{\lambda^3}\Bigr)_{\mu = \mu_0}=
  3.3535\, \Bigl(\frac{v^c}{\lambda^3}\Bigr)_{Bose} \, . $$
 Hence, we can estimate the ratio in our model using
respective estimates from Bose-condensate case~\cite{Harko}.

In Ref.~\cite{Harko}, the BEC concept, using the Gross-Pitaevskii
equation in the Thomas-Fermi approximation, is applied for finding a
space distribution or density profile of the dark matter. It gives
the radius and total mass of the dark matter halo:
\begin{equation}\label{RM}
R=\pi\sqrt{\frac{\hbar^2a}{Gm^3}},\qquad
M=\frac{4}{\pi}R^3\rho^{(c)},\label{MR}
\end{equation}
dependent on the (repulsive) $s$-wave scattering length $a$, the
gravitational constant $G$, mass $m$ of constituent particle and the
central mass density $\rho^{(c)}$.

With varying $R$ and $\rho^{(c)}$, their model is able to reproduce
characteristics of the rotational curves in the outer region of the
galaxies and the total mass $M$. However, it neglects a
supplementary interaction in the inner region and leads to some
overestimation of $M$, compared with the observational data e.g.
those discussed in~\cite{Harko}.

To resolve this, we attempt to account for additional, as compared
to the BEC model, attraction earned due to $\mu$-deformed statistics
and influencing the value $\rho^{(c)}$. Omitting here the dynamics
aspects\footnote{The issues concerning respective modification of
the Gross-Pitaevskii equation and its implications are under study,
and we hope to report on that in near future.} and determining
$\rho^{(c)}$ by the ratio $m/v$, an effective interaction is
included by defining the (critical) volume-per-particle in the case
of $\mu$-deformed thermodynamics: $v=\lambda^3/g^{(\mu)}_{3/2}(1)$.
It means that
$\rho^{(c)}_{(\mu)}=\biggl(g^{(\mu)}_{3/2}(1)/g^{(0)}_{3/2}(1)\biggr)\rho^{(c)}$
 and therefore $M^{(\mu)}=\biggl(g^{(\mu)}_{3/2}(1)/g^{(0)}_{3/2}(1)\biggr)M$
 will play the role of new corrected characteristics. Since
$g^{(\mu)}_{3/2}(1)<g^{(0)}_{3/2}(1)$ at $\mu>0$, these predictions
can give instead of $M^{BEC}\equiv M^{(0)}$ a better agreement with
the observational data.

\section{Concluding remarks}

 Thermodynamics of the $\mu$-Bose gas -- total mean number
 of particles $N^{(\mu)}$, the ($\log$ of) $\mu$-deformed partition function etc.,
 involve $\mu$-generalization $g^{(\mu)}_{k}(z)$ of polylogarithms
$g_{k}(z)$. This fact influences other results in the paper.
 Metric tensor, Christoffel symbols and hence the Riemann curvature
get expressed through $\mu$-polylogarithms. Positive sign of
curvature and its divergence as $z\!\to\!1$ witness the attraction
between $\mu$-particles and confirm Bose-like condensation.
 Formula for the $\mu$-critical temperature $T_c^{(\mu)}$ is
compared with usual Bose case, with infinite statistics system, and
with $T_c^{(p,q)}$ of $p,q$-Bose gas model~\cite{GR-12} (therein,
$T_c^{(p,q)}$ can be both higher and lower, depending on the values
of parameters $p$ and $q$).
 The ratio $T_c^{(\mu)}/{T_c}$ as a function of $\mu$-parameter shows: \
 critical temperature $T_c^{(\mu)}$
 exceeds\footnote{Concerning this property, it would be very
 interesting to give detailed analysis of diverse species
 of cold atoms (Rb, K, Ca etc.),
 for which the existence of BEC
 is already confirmed. However,
 this goes beyond the scope of present paper.}
 critical $T_c$ of usual Bose gas, in contrast with the system of infinite
statistics showing lower critical temperature~\cite{Infinite} than
the usual Bose $T_c$. We consider such property as one of the
virtues, from the viewpoint of applying $\mu$-BGM to modeling dark
matter: in our case stability of the condensate extends higher in
temperature than with usual $T_c$.  Other facts important for the
 modeling and valid for the infinite statistics system~\cite{Infinite},
 e.g. the smallness of particle mass, can be demonstrated for $\mu$-bosons as well.

Further remarkable properties of $\mu$-Bose gas model (say, the
falling behavior of entropy-per-volume versus the $\mu$-parameter --
that means decreasing chaoticity with growing deformation strength),
when used for modeling dark matter, can also lead to interesting
implications.

In the context of dark matter, the inner structure of its
constituents at a given deformation plays a remarkable role in their
response to extrinsic perturbation. The parameter $\mu$ is determined by
specific conditions of the dark matter existence in each galaxy or
a local region of Universe.
   Moreover, since deformed critical temperature obeys the condition
$T^{(\mu)}_c\geq T^{(0)}_c>T$, the present model of dark matter may
be also valid for interstellar environment where the temperature
$T$, concentration of dark matter particles $\varrho$
($T_c\sim\varrho^{2/3}$), and parameter $\mu$ can vary.

   Thus, the $\mu$-Bose gas model has its own virtues and like
   infinite statistics system can be used to effectively model basic
 properties of dark matter.
   In that domain, the present $\mu$-Bose gas model may turn out to be just
   as successful as the ($\tilde{\mu},q$)-deformed analog of Bose gas model
  has shown itself in the effective description, see Fig.~5 in~\cite{GM_NP},
  of the observed (in the STAR/RHIC experiments) non-Bose like
behavior of the intercepts of two-pion correlations. Next, like
in~\cite{GM_NP}, {\it compositeness} of (quasi-)particles which
constitute dark matter may be of importance and can be accounted for
by extending the $\mu$-BGM using the results
of~\cite{GKM2011,GM2012}. Clearly, further study is needed to find
new arguments in favor of the proposed model of dark matter.

\section*{Acknowledgements}
This work was partly supported by The National Academy of Sciences
of Ukraine (project No. 0117U000237), and partly by the Grant
(M.V.Kh.) for Young Scientists of National Academy of Sciences of
Ukraine (No.~0117U003534). We would also like to thank the anonymous
referee(s) for the useful and constructive remarks.



\appendix
\section{Infinite Statistics}

A gas of particles obeying~\cite{Greenberg,Medvedev} infinite
statistics (with parameter $p$) was proposed in \cite{Infinite} as a
model of dark matter.
   Authors considered thermodynamical geometry of this gas,
 and showed with their fig.~1 that condensation does occur
 for such system\footnote{Note that the result of \cite{Infinite} is presented there
as the figure without explicit expressions of calculated metric components, Christoffel
symbols and curvature. For the sake of comparison with the results obtained
in our $\mu$-Bose gas model, we reproduce here explicitly the necessary geometric quantities,
all being expressed in terms of Lerch transcendent~\cite{Lerch}.}.

In \cite{Infinite}, the number of particles and internal energy for
infinite statistics gas are given in integral form as
$N=4Apz\beta^{-d/2-1}I(p z,d/2-1)$ and $U=4Apz\beta^{-d/2-1}I(p
z,d/2)$ respectively, where
$I(\xi,\alpha)=\int_0^{\infty} 
 {{\rm e}^x x^{\alpha}}{(e^{2x}-\xi^2)^{-1}}dx .
$
We express these functions through the Lerch
transcendent~\cite{Lerch}:
\begin{equation}
\Phi(z,s,\alpha) = \frac{1}{\Gamma(s)}
\int\limits_0^{\infty}\frac{t^{s-1}e^{-\alpha t}}{1-ze^{-t}} dt
\end{equation}
hereafter $\Gamma(x)$ is the usual gamma-function.

Let $\xi\equiv pz$, $\Phi_k(\xi)\equiv\Phi\left(\xi^2,d/2+k,1/2\right)$, and $C_k\equiv8(2\beta)^{-d/2-k}\Gamma(d/2+k)$.
The expressions for $N$ and $U$ are rewritten as
\begin{equation}
N=\frac{A}{4\beta}C_0\xi\Phi_0(\xi),\quad
U=\frac{A}{2}C_1\xi\Phi_1(\xi).
\end{equation}

We calculate the metric components in the space of two parameters $(\beta,\gamma)$ in the Fisher-Rao representation,
using the known identity \cite{Lerch} for Lerch transcendent $\Phi(z,s-1,\alpha) = \left(\alpha + z\partial_z\right)
\Phi(z,s,\alpha)$. One has
\begin{equation}
G_{\beta\beta}=AC_1\xi\Phi_1(\xi),\quad
G_{\beta\gamma}=AC_1\xi\Phi_0(\xi),\quad
G_{\gamma\gamma}=AC_0\xi\Phi_{-1}(\xi).
\end{equation}
The Christoffel symbols are found to be $\Gamma_{\beta\beta\beta}=-AC_3\xi\Phi_1(\xi)$,
$\Gamma_{\beta\beta\gamma}=\Gamma_{\gamma\beta\beta}=-AC_2\xi\Phi_0(\xi)$,
$\Gamma_{\beta\gamma\gamma}=\Gamma_{\gamma\beta\gamma}=-AC_1\xi\Phi_{-1}(\xi)$,
$\Gamma_{\gamma\gamma\gamma}=-AC_0\xi\Phi_{-2}(\xi)$.

The resulting expression for the thermodynamical curvature is given as
\begin{equation}\label{Rinf}
R=\frac{4\beta}{A\xi} C_0 C_1 C_2
\frac{-2 \Phi_1 \Phi_{-1}^2+\Phi_0^2
\Phi_{-1}+\Phi_{-2} \Phi_1 \Phi_0}{\left(C_2C_0\Phi_{-1} \Phi_1 -C^2_1\Phi_0^2\right)^2}.
\end{equation}
As it is seen from Fig.~1 in \cite{Infinite} and also follows from
(\ref{Rinf}), in three dimensional space ($d\!=\!3$) sending $zp\to 1$
results in $R \to \infty $ and thus leads to the phase transition
(Bose-like condensation). This fact, and some other properties of
particles with infinite statistics (their mass, weakness of their
interaction etc.) has allowed the authors of \cite{Infinite} to
conclude in favor of ability of such system as possible model of
dark matter.

\end{document}